\documentclass[twocolumn,showpacs,aps]{revtex4}

\usepackage{graphicx}
\usepackage{rotating}

\def\lapprox{{\raise0.5ex\hbox{$<$}\hskip-0.80em\lower0.5ex\hbox{$\sim$}
}}
\def\gapprox{{\raise0.5ex\hbox{$>$}\hskip-0.80em\lower0.5ex\hbox{$\sim$}
}}

\begin{document}
\title{Observation of a Structure in $pp \to pp\gamma\gamma$ near the
  $\pi\pi$ Threshold and its Possible Interpretation by $\gamma\gamma$  
Radiation from Chiral Loops in the Mesonic  $\sigma$ Channel}  
\author{M.~Bashkanov$^1$,
D.~Bogoslawsky$^2$,
H.~Cal\'en$^3$,
F.~Cappellaro$^4$,
H.~Clement$^1$,
L.~Demiroers$^5$,
E. Doroshkevich$^1$,
C.~Ekstr\"om$^3$,
K.~Fransson$^3$,
J.~Greiff$^5$,
L.~Gustafsson$^4$,
B.~H\"oistad$^4$,
G.~Ivanov$^2$,
M.~Jacewicz$^4$,
E.~Jiganov$^2$,
T.~Johansson$^4$,
M.M.~Kaskulov$^1$,
S.~Keleta$^4$,
I.~Koch$^4$,
S.~Kullander$^4$,
A.~Kup\'s\'c$^3$,
A. Kuznetsov$^2$,
P.~Marciniewski$^3$,
R.~Meier$^1$,
B.~Morosov$^2$,
W.~Oelert$^8$,
C. Pauly$^5$,
Y. Petukhov$^2$,
A. Povtorejko$^2$,
R.J.M.Y.~Ruber$^3$,
W.~Scobel$^5$,
B. Shwartz$^{10}$,
T. Skorodko$^1$,
V. Sopov$^{12}$,
J.~Stepaniak$^7$,
V.~Tchernyshev$^{12}$,
P. Th\"orngren Engblom$^4$,
V. Tikhomirov$^2$,
A.~Turowiecki$^{11}$,
G.J.~Wagner$^1$,
U.~Wiedner$^4$,
M.~Wolke$^4$,
A. Yamamoto$^6$,
J.~Zabierowski$^{7}$,
J.~Zlomanczuk$^4$
}
\affiliation{
$^1$Physikalisches Institut der Universit\"at T\"ubingen, D-72076 T\"ubingen, 
Germany\\
$^2$Joint Institute for Nuclear Research, Dubna, Russia \\
$^3$The Svedberg Laboratory, Uppsala, Sweden \\
$^4$Uppsala University, Uppsala,Sweden \\
$^5$Hamburg University, Hamburg, Germany \\
$^6$High Energy Accelerator Research Organization, Tsukuba, Japan \\
$^7$Soltan Institute of Nuclear Studies, Warsaw and Lodz, Poland \\
$^8$Forschungszentrum J\"ulich, Germany \\
$^9$Moscow Engineering Physics Institute, Moscow, Russia \\
$^{10}$Budker Institute of Nuclear Physics, Novosibirsk, Russia \\
$^{11}$Institute of Experimental Physics, Warsaw, Poland \\
$^{12}$Institute of Theoretical and Experimental Physics, Moscow, Russia \\
~~~~~~~~ \\
(CELSIUS-WASA Collaboration)}

\date{\today}

\begin{abstract}
The $pp  \to pp\gamma\gamma$ reaction has been measured at CELSIUS using 
the WASA $4\pi$-detector with
hydrogen pellet target. At $T_p = 1.20$ and 1.36 GeV, where most of the 
statistics has been accumulated, the $\gamma\gamma$ invariant mass spectrum 
exhibits a narrow structure around the 
$\pi\pi$ threshold, which possibly may be
associated with two-photon radiation of $\pi^+\pi^-$ loops in the mesonic 
$\sigma$ channel.
\end{abstract}


\maketitle

The question whether there exists  a low-lying
scalar-isoscalar resonance in the $\pi\pi$ system, the so-called
$\sigma$-meson,  has a long
history. Starting from the $\sigma$ particle problem in nuclear physics, where
such an exchange particle (possibly in form of a correlated $\pi\pi$ exchange)
is needed to accommodate for the scalar-isoscalar attraction in the $NN$
interaction, the quest for the $\sigma$-meson has found renewed interest in
QCD as the chiral partner of the pion --- and in this context possibly even 
as the Higgs particle of the strong interaction \cite{pen}. Also the
scalar-isoscalar 
meson sector is under much debate presently, since there are more states known
meanwhile (including also possible glueball candidates) than can be fitted
into a single multiplet. Hence it has been suggested \cite{clo}, that there are
actually two nonets, a higher-lying one of predominant $q\bar{q}$ nature and a
lower-lying one, encompassing the  $\sigma$ meson, of dominant 
$q\bar{q}q\bar{q}$ structure. This picture of the $\sigma$ would be also 
close to the result of chiral perturbation theory, where the $\sigma$ emerges 
dynamically in $\pi\pi$ rescattering \cite{col,oll,pel}. 

Experimentally the $\sigma$ is very hard to sense because of its expected huge
width due to its fall-apart decay into $\pi\pi$. Besides of $\pi\pi$ phase
shifts recent experiments on $J/\psi \to \omega \pi^+\pi^-$ \cite{nin}, $D^+
\to \pi^-\pi^+\pi^+$ \cite{ait} and $\tau \to \nu_\tau \pi^-\pi^0\pi^0$
\cite{asn} 
have added evidence for the existence of $\sigma$ with a mass in the range
320-480 MeV and a width of roughly the same size. To learn more specifically
about the nature of the $\sigma$ meson, its decay into two gammas would be of
great help, since this would directly give evidence for tight
particle-antiparticle correlations influencing heavily the annihilation
radiation \cite{pdg,boy,mar,bog}. 
In fact, measurements \cite{boy,mar} of double pion production in $e^+e^- \to
\gamma^\ast\gamma^\ast \to \pi\pi$ have been used to get access to the $\sigma
\to \gamma\gamma$ decay \cite{pdg,bog}. At energies near the $\pi\pi$
threshold these reactions are governed by pion production through the
Born terms. In a theoretical analysis of these data a possible contribution of
$\gamma^\ast\gamma^\ast \to \sigma \to \pi\pi$ has been investigated and a
value of $\Gamma_{\sigma\to\gamma\gamma}/\Gamma_{\sigma \to \pi\pi} \approx
10^{-6}$ has been extracted \cite{pdg,bog} at the pole of the $\sigma$.

We have carried out measurements of the reaction $pp \to
pp\gamma\gamma$ at the CELSIUS ring in the energy range of $T_p = 775$ - 1360
MeV using the WASA detector together with the pellet hydrogen target system
\cite{zab}. The detector has nearly full angular coverage for the detection of
charged and uncharged particles. The forward detector consists of a straw
tracker unit followed by plastic scintillator quirl and range hodoscopes,
whereas the central detector comprises in its inner part a thin-walled
superconding magnet containing a minidrift chamber for tracking, and in its
outer part a plastic scintillator barrel surrounded by an electromagnetic
calorimeter consisting of 1012 CsI(Na) crystals. In these runs triggers had
been set to allow also for simultaneous measurements of single pion, double
pion and --- in case of the highest energy --- also $\eta$-production. In case
of $pp \to pp\gamma\gamma$ the two protons have been detected in the forward
detector, whereas the gammas have been detected in the central detector. This
way  the full four-momenta have been measured for all particles, allowing thus
kinematic fits for each event with 4 overconstraints. Also in order to be safe
from events, where particles have escaped detection, we require
the total energy of the detected particles (before kinematic fit) to be equal
to that in the entrance channel within 100 MeV, i.e. within a difference
smaller than the pion mass. To suppress background further we in addition
accept only photons with $E_\gamma > 50$ MeV for the final event samples.

The by far highest statistics has been accumulated at $T_p = 1.36$ GeV,
hence we concentrate in the following on this energy. 
Fig. 1a shows a
scatter plot of the $\gamma\gamma$ invariant mass $M_{\gamma\gamma}$ versus the
pp missing mass $MM_{pp}$ for the selected events. The photon
energy resolution is best for small energies, hence the resolution in
$M_{\gamma\gamma}$ is best at low energies. The situation is reversed in
$MM_{pp}$, since the protons undergo hadronic interactions in the detector
increasingly with increasing energy. Careful inspection of this plot exhibits
already evidence for some enhanced rate between the spots for $\pi^0$ and 
$\eta$.
In order to clean the sample from events,
where the protons have undergone very large hadronic interactions, we employ
the condition $(M_{\gamma\gamma} - MM_{pp})/MM_{pp} < 0.4$. The emerging
 projection of the  scatterplot onto the $M_{\gamma\gamma}$ axis is
shown in Fig. 1b. A peak-like
structure around the $\pi\pi$ threshold appears now clearly
visible. Finally a kinematic fit ( without imposing conditions  for the final 
meson mass ) is applied to the sample, resulting
(Fig. 1c)  primarily  in a  reduction of
the width of the peaks, in particular of the $\eta$ peak. The resulting 
full width at half maximum is $\approx 22$ MeV for $\pi^0$ and $\eta$ peaks. 
The width of the structure at the  $\pi\pi$ threshold also is consistent with 
this value.

An important check whether the peak-like structure  could be of instrumental
origin may be  
provided by measurements under different kinematical conditions. To this end 
we have analyzed in the same way data accumulated at lower energies and 
found evidence for this spike also at $T_p = 1.2$ GeV (Fig. 1d) despite the 
lower statistics there.

\begin{figure}
\resizebox{0.3\textwidth}{!}{%
\includegraphics {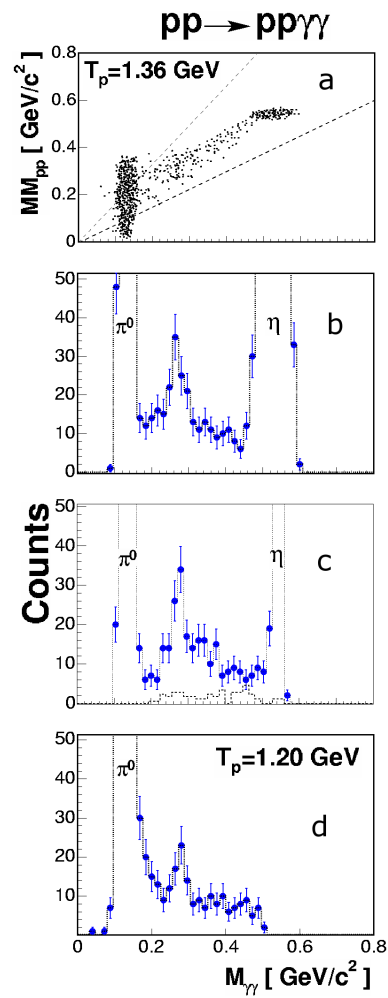}
}
\caption{(a): Scatterplot of 
$M_{\gamma\gamma}$ versus the  $MM_{pp}$ for all selected 
events at $T_p = 1.36$ GeV.  
The dashed line shows the cut 
$(M_{\gamma\gamma} - MM_{pp})/ MM_{pp} < 0.4$ to exclude events, where protons 
have undergone large hadronic interactions in the detector.
(b) - (c): $M_{\gamma\gamma}$ spectrum (projection of the
scatterplot above) before (b) and after (c) kinematic fit.The dotted histogram
 in (c) shows background expected from misidentified $\pi^0\pi^0$ events.
 (d): The same as (c), but for $T_p = 1.2$ GeV. 
}
\end{figure}

This spike turns out to be very stable against cuts. E.g.,
increase of the threshold $E_\gamma = 50$ MeV to $E_\gamma = 100$ MeV has no
significant effect on this spike.  MC simulations for $\pi^0$ and
$\eta$ production providing a  good reproduction of the respective peaks in
$M_{\gamma\gamma}$  give no evidence for a scatter of counts (stemming
from these reactions) into the region between the two peaks. Also 
from simulation of 
$\pi^+\pi^-$ production we do not get any contributions in the 
$M_{\gamma\gamma}$ spectrum. Solely 
the MC simulation of  $\pi^0\pi^0$ production does give some scatter of counts
into this region stemming from $pp\pi^0\pi^0 \to pp\gamma\gamma\gamma\gamma$ 
events, where 
two low-energy $\gamma$s have escaped detection in WASA. In fact, taking into 
account the absolute cross section of $\approx 200 \mu$b for the $\pi^0\pi^0$ 
production at 1.36 GeV 
\cite{bys,koc} about $15 \%$ of the observed counts between 
 $\pi^0$ and $\eta$ peaks are compatible with
misidentified $pp\pi^0\pi^0$ events ( see dotted histogram in fig. 1c ). 
The most dangerous situation for 
producing an artefact peak at $2m_\pi$ is, if two $\pi^0$s decay in such a 
similar manner that the clusters produced by their $\gamma$s just merge 
pairwise into each other. However, such a constellation, which is included in 
the MC simulations shown in fig. 1c, dotted line, is too rare to produce an 
enhancement at $2m_\pi$. We also have investigated this special scenario 
with real 
$\pi^0\pi^0 $ data by deliberately merging clusters. Again this operation did 
not result in a reproduction of the observed structure.

Since none of these simulated 
processes is able to account for the structure observed near the $\pi\pi$
threshold and also detailed and comprehensive tests of detector performance 
and event structures have not given any hint for an artifact, we are led to 
consider seriously the possibility that the observed structure is real and 
might be due to the process 
$pp\to pp\sigma \to pp\gamma\gamma$,
in particular also since $pp \to pp\pi^+\pi^-$ and $pp \to
pp\pi^0\pi^0$ reactions are dominated by $\sigma$ production \cite{bro,pae}.

The statistical significance of the spike observed at the $\pi\pi$
threshold may be estimated directly by inspection of Fig.~1. There we see that
the spike contains roughly $N_S \approx 60$ counts above a background of $N_B
\approx 60$ counts at $T_p = 1.36$ GeV. The 
significance in
standard deviations depends on the assumption about the background
behaviour. Neglect of the statistical fluctuations of the background gives $S
= N_S/\sqrt{N_B} \approx 7.7$. 
A more reliable estimate is $S = N_S/\sqrt{N_S+N_B} \approx 5.5$, assuming a
fluctuating background, which is smooth and well-fixed in shape. Finally,
assuming the full uncertainty of a statistically independent background
results in $S = N_S/\sqrt{N_S+2*N_B} \approx 4.5$. 
Somewhat lower values arise, if we use the data at 
 $T_p = 1.2$ GeV, where $N_S \approx N_B \approx 30$. 

We note that recently the $\sigma \to \gamma\gamma$ decay has also been 
searched for \cite{sta} in $^{12}C(\pi^-, \pi^0\pi^0)$. The 
$M_{\gamma\gamma}$ spectrum shown there exhibits some 
indication of a gentle enhancement near $2m_\pi$ . Though this reaction cannot 
directly be compared to our case, it is worth noting that the quoted relative 
upper limit is of similar order of magnitude as our observation.

\begin{figure}
\resizebox{0.30\textwidth}{!}{%
\includegraphics {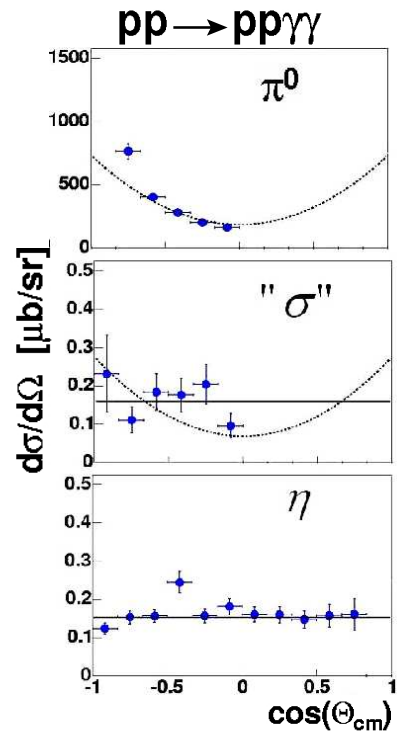}
}
\caption{ Angular distributions of the $\gamma\gamma$ pair momentum in
  the overall cms for $\gamma$s belonging to $\pi^0$ peak,
  $\eta$ peak and to the spike region,
  respectively. The solid lines show fits with $\sigma(\Theta_{CM}) \sim 1 +
  3\cos^2 \Theta_{CM}$ for $\pi^0$ and $\sigma(\Theta_{CM}) = $ constant
  otherwise.\
}
\end{figure}

We next inspect the efficiency and acceptance corrected angular distributions. 
 For these we reconstruct the $\pi^0$, $\eta$ and $\sigma$ momenta from the 
respective $\gamma$ pairs. The angular distributions
$\sigma(\Theta_{CM})$ of these reconstructed momenta in the overall center of
mass system are plotted in Fig. 2. For $\pi^0$ we obtain an angular
distribution 
close to $(1+ 3 \cos^2 \Theta_{CM}$) consistent with $\Delta$ and Roper 
resonance excitations in
$pp \to pp\pi^0$. For $\eta$ we see an
s-wave behavior consistent with the excitation of the $S_{11}(1535)$ resonance
in $pp \to pp\eta$ and also in agreement with previous measurements
\cite{cal,abd}. And for the spike region  we obtain a flat 
angular distribution, too,  again consistent with s-wave production.


In order to get absolute cross sections we have for convenience normalized our
data for the $\eta$ peak to the well-known cross section of $4.9 \mu$b for $pp
\to pp\eta$ at $T_p = 1.36$ GeV \cite{cal}, which by use of the known $\eta
\to \gamma\gamma$ branching ratio of 39\% \cite{pdg} results in a cross
section of 1.9 $\mu$b for $pp \to pp\eta \to pp\gamma\gamma$. This 
normalization then leads to a cross section 
value of 4 mb for $pp \to pp\pi^0$ which is in reasonable agreement with
previous measurements \cite{bys}. For the structure around the $\pi\pi$
threshold the quotation of an absolute cross section depends crucially on the
assumption of background beneath this structure. If we just take 
the narrow spike above the smooth continuum, then we arrive at
roughly $1 - 2 \mu$b for its cross section. The
acceptance and efficiency corrected as well as normalized $M_{\gamma\gamma}$
spectrum is shown in Fig. 3.

\begin{figure}
\resizebox{0.30\textwidth}{!}{%

\includegraphics {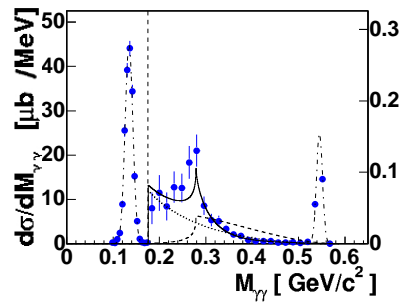}
}
\caption{ Differential cross section in dependence of
$M_{\gamma\gamma}$. Note the change of scale at 
$M_{\gamma\gamma}=0.18 GeV/c^2$ by more than a factor of 100! The dotted 
curve shows MC simulations for double
bremsstrahlung adjusted in height to the data. The
dashed line gives calculations of the diagrams in Fig. 4, whereas the
solid line represents their coherent sum.}

\end{figure}

\begin{figure}
\resizebox{0.25\textwidth}{!}{%
\includegraphics {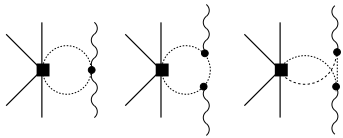}
}
\caption{ Graphs for the dynamical formation of s-wave virtual 
$\pi^+\pi^-$ loops generated in the $pp$
collision process and its subsequent annihilation into two photons.
}
\end{figure}

A process which principally contributes to the energy region of interest
is double bremsstrahlung. Single bremsstrahlung 
has been calculated \cite{shk} at $T_p = 1.35$ GeV to give a total cross 
section of about 100$\mu$b for $E_\gamma > 100$ MeV. For double bremsstrahlung
we are not aware of any such detailed calculation. However, we may obtain a 
first crude estimate for its cross section by just multiplying the single 
bremsstrahlung result with the fine structure constant $\alpha$. This estimate
presumably  will only be a lower limit, since there are much more different 
diagrams contributing to  
double than to single bremsstrahlung. Anyway from these considerations we may 
expect the double bremsstrahlung cross section to be of the same order of 
magnitude as  given by the observed  continuum  of counts between $\pi^0$ and 
$\eta$ peaks. Though double bremsstrahlung certainly cannot explain the narrow 
spike, it appears to be a good candidate for explaining the  continuum beneath
the spike. The dotted lines in Fig. 3 shows a MC simulation for
this process adjusted in  height to the data and assuming
$d\sigma/dM_{\gamma\gamma} \sim M_{\gamma\gamma}^{-1}$  \cite{smi}.

Qualitatively a cusp-like energy dependence of the process 
$pp\to pp\sigma \to pp\gamma\gamma$ can be obtained by a Breit-Wigner ansatz 
$d\sigma/d~M_{\gamma\gamma} \sim (\varepsilon^2 +
\Gamma^2/4)^{-1}$ with $\varepsilon = M_{\gamma\gamma} - m_\sigma$ and the 
( energy dependent ) total width 
$\Gamma = \Gamma_{\gamma\gamma} + \Gamma_{\pi\pi}$. Below the $\pi\pi$ 
threshold we have 
$\Gamma = \Gamma_{\gamma\gamma} \approx 10^{-6} \Gamma_{\pi\pi}$ and above 
$\Gamma \approx \Gamma_{\pi\pi}$. For 
$m_\sigma \approx \Gamma_{\pi\pi} \approx 350 MeV$ we this way arrive at
a qualitative description of the observed structure.

To discuss the possible origin of the spike on a more fundamental level, we 
consider the graphs in Fig. 4, where correlations in the $\sigma$ channel are 
formed dynamically by pion loops generated in the pp collision process  and 
coupled to the $\gamma\gamma$ channel. The blob represents the generation of 
virtual pion pairs and may be adjusted in scale to the total hadronic pp 
cross section 
at $T_p = 1.36$ GeV. Evaluation of the loop  
diagrams with a branch cut at the $\pi^+\pi^-$ threshold \cite{kas},  leads 
to a cusp-like structure (dashed lines in Fig. 3) with a strength  
basically of order $\alpha^2$ smaller than the total cross section. Though we 
arrive at the proper order of magnitude for the effect under discussion, the 
predicted shape is still quite different. 
However,  this amplitude is to interfere with the 
one for double bremsstrahlung. With a proper choice for the relative 
phase between both processes we obtain a curve, which fits the data remarkably 
well (solid lines in Fig. 3). Note that this cusp effect due to 
$\gamma\gamma$ radiation induced by recombination of chiral loops is very 
general and should appear principially also in channels other than the 
$\sigma$ channel discussed here.

Summarizing we observe a narrow spike around the $\pi\pi$ threshold in 
the $M_{\gamma\gamma}$ spectrum of the  $pp  \to pp\gamma\gamma$ reaction. Its 
angular distribution is consistent with s-wave production. In a first 
attempt 
for its interpretation we consider $\sigma$ channel pion loops , which are 
generated by the pp collision process and which decay into the $\gamma\gamma$
 channel. Chiral loop  calculations of this process reveal indeed a 
cusp-like behaviour, which by interference with the underlying double 
bremsstrahlung 
background can give a reasonable account of the data.
If true then this observation would be a manifestation of a correlated
$\pi\pi$ system in the $\sigma$ channel. The correlated $\pi\pi$ exchange
between nucleons \cite{ose,mmk} would not need to be lifted above the 
$\pi\pi$ threshold to
get on mass shell, it would be so already below this threshold with respect
to the $\gamma\gamma$ decay channel. Similar loop effects are expected to show 
up also in other channels.


We are grateful to the TSL/ISV personnel for the continued help during the 
course of these measurements. This work has been supported by BMBF (06TU201), 
DFG (Europ. Graduiertenkolleg 683) and Baden-Wuerttemberg 
Landesforschungsschwerpunkt.

\end{document}